 \documentclass[final,5p,times,twocolumn,12pt,sort&compress]{elsarticle}
\usepackage{soul}
\usepackage{lineno,hyperref}
\usepackage{txfonts}
\usepackage{graphicx}
\usepackage{dcolumn}% Align table columns on decimal point
\usepackage{array}
\usepackage{color}
\usepackage{CJK}
\usepackage{multicol}
\usepackage{amssymb}
\usepackage{amsmath}
\usepackage{type1cm}
\usepackage{caption}
\usepackage{natbib}
\usepackage{lineno,hyperref}
\usepackage{comment}
\usepackage{longtable}
\usepackage{ctable}
\modulolinenumbers[5]

\def\kw#1{{\color{black} #1}}

\onecolumn

\journal{JQSRT, prepared by using elsarticle.cls}
\begin{document}
\begin{frontmatter}
\title{Extended calculations of energy levels, radiative properties,  and lifetimes for
oxygen-like \mbox{Mo XXXV}}

\author[hb,fd]{Kai Wang}
\author[hb]{Wei Zheng}
%\author[fd]{C. Y. Zhang}
\author[hb]{Xiao Hui Zhao\corref{zhao}}
\author[nudt]{Zhan Bin Chen\corref{bin}}
\author[fd]{Chong Yang Chen}
\author[bj]{Jun Yan}

\cortext[zhao]{xhzhao@hbu.edu.cn}
\cortext[bin]{chenzb008@qq.com}
\address[hb]{Hebei Key Lab of Optic-electronic Information and Materials, The College of Physics Science and Technology, Hebei University, Baoding 071002, China}
\address[fd]{Shanghai EBIT Lab, Key Laboratory of Nuclear Physics and Ion-beam Application, Institute of Modern Physics, Department of Nuclear Science and Technology, Fudan University, Shanghai 200433, China}
\address[nudt]{School of Science, Hunan University of Technology, Zhuzhou, 412007, China}
\address[bj]{Institute of Applied Physics and Computational Mathematics, Beijing 100088, China}

\begin{abstract}
Employing two state-of-the-art methods,  second-order many-body perturbation theory  and multiconfiguration Dirac-Fock, highly accurate calculations are performed for the lowest 318 fine-structure levels arising from the $2s^{2} 2p^{4}$, $2s 2p^{5}$, $2p^{6}$, $2s^{2} 2p^{3} 3l$, $2s 2p^{4} 3l$, $2p^{5} 3l$, and $2s^{2} 2p^{3} 4l$ configurations in O-like \mbox{Mo XXXV}. 
Complete and consistent atomic data, including excitation energies, lifetimes,  wavelengths, and  E1, E2, M1 line strengths,  oscillator strengths, and transition rates  among these 318 levels are provided. 
Comparisons are made between the present two data sets, as well as with other available experimental and theoretical values. 
The present data are accurate enough for identification and deblending of emission lines involving the $n=3,4$ levels and are also useful for modeling and diagnosing fusion plasmas\kw{. These data can} be considered as a benchmark for other calculations.
\end{abstract}


\end{frontmatter}

\twocolumn
\section{Introduction}
Walls of fusion reactors often contain alloys of molybdenum, and ions of Mo are present in the plasmas due to sputtering from the walls~\cite{Mansfield.1978.V11.p1521,Reader.2015.V48.p144001}. Therefore, in order to simulate and diagnose plasmas that contain Mo as a constituent, accurate atomic data for different Mo ions are required. In view of this, soft X-ray emission lines from molybdenum plasmas generated by dual laser pulses were measured for different  Mo ions~\cite{Lokasani.2016.V109.p194103}. ~\citet{Feldman.1991.V8.p531} identified 13 lines of the $(1s^2)2s^22p^4$ - $2s2p^5$ transitions in a laser-produced plasma.

\kw{On} the theoretical side, excitation energies and radiative transition data for the low-lying states of the $2s^22p^4$, $2s2p^5$, and $2p^6$  configurations were provided by different calculations~\cite{Fontes.2015.V101.p143,Hu.2011.V9.p1228,Gu.2005.V89.p267,Zhang.2002.V82.p357,Vilkas.1999.V60.p2808}.  Atomic parameters of the $n > 2$ levels are also needed for applications in plasma physics~\citep{Rice.2000.V33.p5435,Kink.2001.V63.p46409}.

The present study is to provide a complete accurate data set of energy levels,  radiative transition data, and lifetimes involving high-lying levels \kw{with principal quantum numbers $n=3,4$} in O-like \mbox{Mo XXXV}.  \kw{This study is an extension} of our previous work \kw{on} O-like ions~\cite{Wang.2017.V229.p37,Wang.2017.V194.p108}.  
Excitation energies, wavelengths, line strengths, oscillator strengths, transition rates, and lifetimes are provided \kw{here for} the lowest 318 levels of the $2s^{2} 2p^{4}$, $2s 2p^{5}$, $2p^{6}$, $2s^{2} 2p^{3} 3l$ ($l=s, p, d$), $2s 2p^{4} 3l$ ($l=s, p, d$), $2p^{5} 3l$ ($l=s, p, d$), and $2s^{2} 2p^{3} 4l$ ($l=s, p, d, f$) configurations using the many-body perturbation theory (MBPT) method~\citep{Lindgren.1974.V7.p2441,Safronova.1996.V53.p4036,Vilkas.1999.V60.p2808,Gu.2005.V156.p105,Gu.2007.V169.p154}. In order to assess the accuracy of our MBPT calculations, the multiconfiguration Dirac-Fock (MCDF) and relativistic configuration interaction (RCI) method \citep{Grant.2007.V.p,FroeseFischer.2016.V49.p182004} is used to calculate the corresponding data. 
The present study significantly increases the amount of accurate data for the $n = 3, 4$ levels of O-like Mo\kw{.  The accuracy of our results is sufficient} to directly aid and confirm experimental identifications. 

\section{Calculations}
The MBPT method integrated in the FAC code~\cite{Gu.2008.V86.p675}  and the MCDF  method implemented in the GRASP2K code~\cite{Jonsson.2007.V177.p597,Jonsson.2013.V184.p2197} are used to perform the calculations. Both of the methods have been successfully used to calculate  atomic parameters for L- and M-shells systems with high accuracy~\citep{Wang.2014.V215.p26,Wang.2015.V218.p16,Wang.2016.V223.p3,Wang.2016.V226.p14,Wang.2017.V119.p189301,Wang.2017.V194.p108,Wang.2017.V187.p375,Wang.2017.V229.p37,Wang.2018.V235.p27,Wang.2018.V239.p30,Wang.2018.V234.p40,Wang.2018.V208.p134,Wang.2018.V864.p127,Wang.2018.V220.p5,Chen.2017.V113.p258,Chen.2018.V206.p213,Chen.2019.V234.p90,Chen.2019.V225.p76,Guo.2015.V48.p144020,Guo.2016.V93.p12513,Si.2016.V227.p16,Si.2017.V189.p249,Si.2018.V239.p3,Zhao.2018.V119.p314}. We only give an outline of the MBPT and MCDF calculations, since these two methods are described in our earlier work in detail.

\subsection{MBPT}
In the MBPT method, the Hilbert space of the system is divided into two subspaces, including a model space $M$ and an orthogonal space $N$.
By means of solving the eigenvalue problem of a non--Hermitian effective Hamiltonian in the space $M$, we can get the true eigenvalues of the Dirac--Coulomb--Breit Hamiltonian.
The configuration interaction effects in the $M$ space is exactly considered, and the interaction of the spaces $M$ and $N$ is accounted for with the many-body perturbation theory up to the second order.
In the  calculations, we include all states of the $2s^{2} 2p^{4}$, $2s 2p^{5}$, $2p^{6}$, $2s^{2} 2p^{3} 3l$ ($l=s, p, d$), $2s 2p^{4} 3l$ ($l=s, p, d$), $2p^{5} 3l$ ($l=s, p, d$), $2s^{2} 2p^{3} 4l$ ($l=s, p, d, f$), and $2s 2p^{4} 4s$ configurations in the model space $M$,
Through the single and double (SD) virtual excitations of the states spanning the $M$ space, \kw{all interacting virtual states necessary to properly describe inter-electron correlations} are contained in the space
$N$. The maximum $n$ values for the single/double excitations are 200/65, respectively, while the maximum $l$ value is~20. The leading quantum electrodynamic (QED) effects are also considered in our work.

\subsection{MCDF}
The MCDF method has been described by~\citet{FroeseFischer.2016.V49.p182004}.  
Based on the active space (AS) approach~\cite{Sturesson.2007.V177.p539} for the generation of the configuration state function (CSF) expansions, separate MCDF calculations are done for the even and odd parity states. We start the calculation without any excitation from the reference configurations which is usually referred to as the Dirac-Fock (DF) calculation. 
The reference configurations are \kw{the same as those included in the $M$ space of the MBPT calculations described above}. 
Subsequently, the CSFs expansions are obtained through the SD excitations from the  shells of the reference configurations up to a $ n_{max}l_{max} $ orbital, with $n_{max} \leq 8$ and $l_{max} \leq 5$. 
To reduce the number of CSFs, the $1s^2$ core is closed during the relativistic self-consistent field (RSCF) calculations, but is opened during the RCI calculations, where the Breit interaction and  QED effects , i.e., vacuum polarization and self- energy, are included in the Hamiltonian. 
The number of CSFs in the final even and odd state expansions are
approximately 5~060~000  and 3~930~000, respectively, distributed over the different $J$ symmetries.

To provide the $LSJ$ labeling system used in databases such as the Atomic Spectra Database (ASD) of the National Institute of Standards and Technology (NIST)~\cite{Kramida.2018.V.p}, the wave functions in both the MCDF and MBPT calculations  are
transformed  from a $jj$-coupled CSF basis into a
$LSJ$-coupled CSF basis using the methods developed by Gaigalas \citep{Gaigalas.2004.V157.p239,Gaigalas.2017.V5.p6}.

\section{Results and Discussions}
The computed excitation energies for the lowest 318 levels of the $2s^{2} 2p^{4}$, $2s 2p^{5}$, $2p^{6}$, $2s^{2} 2p^{3} 3l$ ($l=s, p, d$), $2s 2p^{4} 3l$ ($l=s, p, d$), $2p^{5} 3l$ ($l=s, p, d$), and $2s^{2} 2p^{3} 4l$ ($l=s, p, d, f$) configurations from our MBPT and MCDF calculations are listed in Table~\ref{table1}, along with the radiative lifetimes estimated from E1, E2, and M1 transition rates, and the $LSJ$-coupled and $jj$-coupled labels obtained from our calculations. 
Table~\ref{table2} lists wavelengths $\lambda_{ij}$, and E1, E2, M1 line strengths $S_{ji}$, oscillator strengths $g_{i}f_{ji}$, and radiative rates $A_{ji}$ among the 318 energy levels along with branching fractions (${\rm BF}_{ji} = \frac{A_{ji}}{\sum \limits_{k=1}^{j-1} A_{jk}}$), obtained from both the MBPT and MCDF methods. All the E1 and E2 values are computed in the Babushkin gauge (equivalent to the non-relativistic length form), which is considered to be more accurate than the Coulomb gauge (equivalent to the non-relativistic velocity form).

\subsection{Excitation energies}
Since excitation energies for O-like Mo are only available for the $n=2$ levels in the previous experimental and theoretical studies, the MBPT and MCDF excitation energies are compared with the experimental values from the NIST ASD, and the other theoretical results calculated by \citet{Gu.2005.V89.p267} and \citet{Vilkas.1999.V60.p2808} in Table~\ref{table3}. 
The average difference with the standard deviation \kw{from} the NIST values are $-249 \pm 748$ cm$^{-1}$ for MBPT, $-68 \pm 471$ cm$^{-1}$ for MCDF, $1053 \pm 707$ cm$^{-1}$ for Gu, and $1955 \pm 1758$ cm$^{-1}$ for \kw{Vilkas et al.}, respectively. Comparing with the previous calculations, there is generally a better agreement between the NIST values and our MBPT (MCDF) results due to \kw{the larger extent of the} electron correlation effects included in our work.

For the remaining levels belonging to the $n = 3$ and $n = 4$ configurations, the relative difference between our two calculations for each level is shown in Table~\ref{table1}. The average absolute difference with standard deviation of the present MBPT and MCDF energy values are $296\pm646$ cm$^{-1}$, corresponding to the average relative difference with the standard deviation of $0.001~\%\pm0.002~\%$, which are satisfactory.

\subsection{Wavelengths, transition rates, and lifetimes}\label{sec_tr}
~\citet{Feldman.1991.V8.p531} identified 13 lines of the $2s^22p^4$ - $2s2p^5$ transitions in a laser-produced plasma, and these lines have been compiled in the NIST ASD. In Table~\ref{table4}, the experimental wavelengths for these 13 lines are compared with the present MBPT and MCDF results. Transition rates from our  MBPT and MCDF calculations are also compared with each other in the table. As \kw{shown} in Table~\ref{table4}, the difference between the NIST and MBPT(MCDF) wavelength for each transition is within 0.1~\%, which is \kw{quite} satisfactory. Our MBPT and MCDF transition rates show good agreement with each other, the \kw{differences being} within 5~\%.

\kw{In the spirit of} the uncertainty estimation method suggested by~\citet{Kramida.2013.V63.p313,Kramida.2014.V2.p22,Kramida.2014.V212.p11}, the difference $\delta S$ of line strengths $S$ from the MBPT and MCDF calculations for each E1 transition is defined as $\delta S$ = $\left|S_{\rm MCDF}  - S_{\rm MBPT} \right|$/max($S_{\rm MCDF}$,~$S_{\rm MBPT}$). The averaged uncertainties $\delta S_{av}$ for line strengths $S$  from the present MBPT and MCDF calculations for E1 transitions in various ranges of $S$ are assessed to 1.7~\% for $S \geq 10^{-1}$; 4.1~\% for $10^{-1} > S \geq 10^{-2}$; 6.1~\% for $10^{-2} > S \geq 10^{-3}$; 8.7~\% for $10^{-3} > S \geq 10^{-4}$; 18~\% for $10^{-4} > S \geq 10^{-5}$, and 35~\% for $10^{-5} > S \geq 10^{-6}$. Then, the larger of $\delta S_{ij}$ and $\delta S_{av}$ is accepted as the uncertainty of each particular line strength. About 1.4~\% have uncertainties of  $\leq$ 2~\% (the category A+), 1.0~\% have uncertainties of  $\leq$ 3~\% (the category A), 20.0~\% have uncertainties of  $\leq$ 7~\% (the category B+), 23.3~\% have uncertainties of  $\leq$ 10~\% (the category B), 8.9~\% have uncertainties of  $\leq$ 18~\% (the category C+), 19.6~\% have uncertainties of  $\leq$ 25~\% (the category C), and 13.5~\% have uncertainties of  $\leq$ 40~\% (the category D+), while  12.3~\% have uncertainties of  $>$ 40~\% (categories D and E). 

Again, using the method suggested in~\cite{Kramida.2014.V2.p22,Kramida.2014.V212.p11}, the uncertainties of the $S$ values for M1 and E2 transitions  are estimated. The estimated uncertainties for all E1, M1, and E2 transitions with BF  $\geq 10^{-5}$ are listed in Table~\ref{table2}.

Our MBPT and MCDF lifetimes for the lowest 318  levels of the $n \leq 4$ configurations in O-like \mbox{Mo XXXV}, which are calculated by considering all possible E1, M1, and E2 transitions, are listed in Table~\ref{table1}. For the lowest 10 levels of the $n = 2$ complex, the MCDF lifetimes agree very well with the MBPT results, and the differences are within 3~\%.  For the remaining 308 levels belonging to the $n = 3$ and $n = 4$ configurations, the average difference with the standard deviation  \kw{is} $0.5~\% \pm 4.2~\%$.

\section{Conclusions}
Employing two state-of-the-art methods (MBPT and MCDF),  excitation energies and lifetimes of the lowest 318 levels for the $n \leq 4$ configurations have been calculated for O-like \mbox{Mo XXXV}. Wavelengths, E1, M1, and E2  transition rates, line strengths, and
oscillator strengths for the transitions among these 318 levels are also reported. 

The accuracy of energy levels and transition probabilities is estimated by comparing the MBPT and MCDF results with available experimental and theoretical data. The average difference with the standard deviation with the NIST values for the \kw{$n=2$} levels are $-249 \pm 748$ cm$^{-1}$ for MBPT \kw{and} $-68 \pm 471$ cm$^{-1}$ \kw{for MCDF,} which indicates the high accuracy of the present two calculations. 
For the $n=3, 4$ levels, the average absolute difference of the present MBPT and MCDF energy values \kw{is} $296\pm646$ cm$^{-1}$, corresponding to the average relative difference  of $0.001~\%\pm0.002~\%$, \kw{where the standard deviations are indicated after the values.} 
The uncertainty of \kw{the}  line strength  is assessed  for each transition, and is available in Table~\ref{table2}. 
The present calculations provide  a consistent and accurate data set for line identification and modeling purposes, which can also be considered as a benchmark for other calculations.

\section{Acknowledges}
We acknowledge the support from the National Key Research and Development Program of China under Grant No.~2017YFA0403200,  the Science Challenge Project of
China Academy of Engineering Physics (CAEP) under
Grant No.~TZ2016005, the National Natural Science Foundation of China (Grant No.~11703004, No.~11674066, No.~11504421, and No.~11474034), the Natural Science Foundation of Hebei Province, China (A2019201300 and A2017201165), and the Natural Science Foundation of Educational Department of Hebei Province, China (BJ2018058). 
Kai Wang expresses his gratitude to the support from the visiting researcher program of  Fudan University.

\onecolumn

%\section*{References}
\bibliography{ref.bib}
%\bibliography{../../../../../article/ref.bib}

\onecolumn

\clearpage

%\section*{Tables}
\newpage
\linespread{1}
\scriptsize
\setlength{\tabcolsep}{5pt}
\begin{longtable}{cllrrrccl}
	\caption{\label{table1}Excitation energies ($E$ in cm$^{-1}$) and lifetimes ($\tau$ in s$^{-1}$) for the lowest  318 levels arising from the $2s^{2} 2p^{4}$, $2s 2p^{5}$, $2p^{6}$, $2s^{2} 2p^{3} 3l$, $2s 2p^{4} 3l$, $2p^{5} 3l$, and $2s^{2} 2p^{3} 4l$ configurations of O-like \mbox{Mo XXXV} from the present MBPT and MCDF calculations. $J\pi$: the total quantum number $J$ and parity $\pi$; $\Delta E$ -- the differences  (in cm$^{-1}$)  between the MBPT and MCDF calculated  excitation energies. }\\
	\toprule
	Key & $LSJ$-coupled CSF& $J\pi$ &$E_{\rm MCDF}$  & $E_{\rm MBPT}$& $\Delta E$ & $\tau_{\rm MCDF}$ & $\tau_{\rm MBPT}$ & $jj$-coupled CSF$^{a,b,c}$\\
	\midrule
	\endfirsthead
	\caption{(\emph{continued})} \\
	\toprule
	Key & $LSJ$-coupled CSF& $J\pi$ &$E_{\rm MCDF}$  & $E_{\rm MBPT}$& $\Delta E$ & $\tau_{\rm MCDF}$ & $\tau_{\rm MBPT}$ & $jj$-coupled CSF$^{a,b,c}$\\
		\midrule
	\endhead
	\endfoot
	\bottomrule
	\endlastfoot
1    &$2s^{2}\,2p^{4}(^{3}P)~^{3}P_{2}$                             &2+   &0              &0         &0         &               &               &2p+2(4)4                                               \\
2    &$2s^{2}\,2p^{4}(^{1}S)~^{1}S_{0}$                             &0+   &207772         &207830    &57        &7.987E-02      &7.987E-02      &2p+2(0)0                                               \\
3    &$2s^{2}\,2p^{4}(^{3}P)~^{3}P_{1}$                             &1+   &863686         &863215    &-471      &7.767E-08      &7.767E-08      &2p-1(1)1.2p+3(3)2                                      \\
4    &$2s^{2}\,2p^{4}(^{1}D)~^{1}D_{2}$                             &2+   &976768         &976432    &-337      &1.414E-07      &1.414E-07      &2p-1(1)1.2p+3(3)4                                      \\
5    &$2s^{2}\,2p^{4}(^{3}P)~^{3}P_{0}$                             &0+   &1921492        &1920667   &-825      &2.842E-08      &2.842E-08      &2p+4(0)0                                               \\
6    &$2s~^{2}S\,2p^{5}~^{3}P_{2}^{\circ}$                          &2-   &2379704        &2380355   &651       &4.938E-12      &4.938E-12      &2s+1(1)1.2p+3(3)4                                      \\
7    &$2s~^{2}S\,2p^{5}~^{3}P_{1}^{\circ}$                          &1-   &2654158        &2655091   &933       &2.294E-12      &2.294E-12      &2s+1(1)1.2p+3(3)2                                      \\
8    &$2s~^{2}S\,2p^{5}~^{3}P_{0}^{\circ}$                          &0-   &3289601        &3289830   &228       &3.061E-12      &3.061E-12      &2s+1(1)1.2p-1(1)0                                      \\
\end{longtable}
\begin{enumerate}[]
	\item $^a$ The number at the end or  \kw{in parentheses} is 2$J$.
	\item $^b$ $s+ = s_{1/2}$, $p- = p_{1/2}$, $p+ = p_{3/2}$, $d- = d_{3/2}$, $d+ = d_{5/2}$, $f- = f_{5/2}$,  and $f+ = f_{7/2}$.
	\item $^c$ The number after $\pm$ is the occupation number of the corresponding sub-shell. For example, the $jj$-coupled CSF of level 9 is $2s_{1/2} 2p_{1/2}2p_{3/2}^{4}$. The full sub-shell $2p_{3/2}$ is omitted.
	\end{enumerate}
\clearpage

\clearpage
\linespread{1}
\tiny
\setlength{\tabcolsep}{5pt}
\begin{longtable}{llcccccccccccc}
	\caption{\label{table2}Wavelengths ($\lambda$, in vacuum \AA), transition rates ($A$, in s$^{-1}$), oscillator strengths ($gf$, dimensionless), line strengths ($S$, in a.u.), and branching fractions (BF) for the transitions among the 318 levels  listed in Table~\ref{table1} from the present MBPT and MCDF calculations. The two first columns ($i$ and $j$) refer to the keys of Table~\ref{table1}. The last column \kw{represents} the estimated accuracies of the $S$ values using \kw{the terminology} of the NIST ASD~\citep{Kramida.2018.V.p}. Only transitions with BF  $\geq 10^{-5}$ are presented.}\\
	\toprule
	$i$ & $j$& Type & $\lambda_{\rm MBPT}$ & $A_{\rm MBPT}$ & $gf_{\rm MBPT}$ & $S_{\rm MBPT}$ & $\rm BF_{\rm MBPT}$& $\lambda_{\rm MCDF}$ & $A_{\rm MCDF}$ & $gf_{\rm MCDF}$ & $S_{\rm MCDF}$ & $\rm BF_{\rm MCDF}$& Acc.\\	\midrule
	\endfirsthead
	\caption{(\emph{continued})} \\
	\toprule
	$i$ & $j$ Type & $\lambda_{\rm MBPT}$ & $A_{\rm MBPT}$ & $gf_{\rm MBPT}$ & $S_{\rm MBPT}$ & $\rm BF_{\rm MBPT}$& $\lambda_{\rm MCDF}$ & $A_{\rm MCDF}$ & $gf_{\rm MCDF}$ & $S_{\rm MCDF}$ & $\rm BF_{\rm MCDF}$& Acc.\\

\midrule
	\endhead
	\midrule
	\endfoot
	\bottomrule
	\endlastfoot
    1  &  2&  E2& 4.81296E+02&    1.252E+01&  4.348E-10&  2.887E-04&  1.000E+00& 4.81163E+02&  1.243E+01&  4.314E-10&  2.862E-04 & 1.000E+00&    B+\\
1  &  3&  M1& 1.15783E+02&    1.066E+07&  6.430E-05&  1.841E+00&  8.279E-01& 1.15846E+02&  1.066E+07&  6.435E-05&  1.844E+00 & 8.280E-01&    AA\\
1  &  3&  E2& 1.15783E+02&    4.675E+03&  2.819E-08&  2.605E-04&  3.631E-04& 1.15846E+02&  4.610E+03&  2.782E-08&  2.576E-04 & 3.581E-04&    B+\\
2  &  3&  M1& 1.52459E+02&    2.211E+06&  2.311E-05&  8.713E-01&  1.717E-01& 1.52582E+02&  2.209E+06&  2.313E-05&  8.728E-01 & 1.716E-01&    AA\\
1  &  4&  M1& 1.02378E+02&    7.057E+06&  5.545E-05&  1.404E+00&  9.979E-01& 1.02414E+02&  7.056E+06&  5.548E-05&  1.405E+00 & 9.977E-01&    AA\\
1  &  4&  E2& 1.02378E+02&    9.136E+03&  7.180E-08&  4.587E-04&  1.292E-03& 1.02414E+02&  9.036E+03&  7.105E-08&  4.545E-04 & 1.278E-03&    B+\\
2  &  4&  E2& 1.30040E+02&    8.323E+02&  1.055E-08&  1.382E-04&  1.177E-04& 1.30106E+02&  8.266E+02&  1.049E-08&  1.376E-04 & 1.169E-04&    B+\\
3  &  4&  M1& 8.84311E+02&    5.022E+03&  2.944E-06&  6.438E-01&  7.101E-04& 8.83264E+02&  5.043E+03&  2.949E-06&  6.442E-01 & 7.131E-04&    AA\\
\end{longtable}
\begin{enumerate}[]
	\item Only transitions among the lowest 10 levels of the $n=2$ configurations are shown here. Table~\ref{table2} is available online in its entirety \kw{at} the \emph{JQSRT} website.
\end{enumerate}
\clearpage

\newpage
\linespread{1}
\scriptsize
\setlength{\tabcolsep}{5pt}
\begin{longtable}{llccccccccc}
	\caption{\label{table3}Energy levels (in cm$^{-1}$) relative to the ground state for the $n=2$ states. NIST -- the experimental values from the NIST ASD~\citep{Kramida.2018.V.p}; MBPT -- the present MBPT results; MCDF -- the present MCDF results; Gu -- the theoretical values from \citet{Gu.2005.V89.p267};  Vilkas -- the theoretical values from \citet{Vilkas.1999.V60.p2808}. $\Delta E$ -- the differences  (in cm$^{-1}$)  between calculated energies from different calculations and the NIST values.}\\
    \midrule
    Key & State & NIST & \multicolumn{2}{c}{MBPT} &  \multicolumn{2}{c}{MCDF} & \multicolumn{2}{c}{Gu} & \multicolumn{2}{c}{Vilkas}\\
    \cmidrule(r){4-5}\cmidrule(r){6-7}\cmidrule(r){8-9}\cmidrule(r){10-11}
        &       &   $E$  &  $E$                       &  $\Delta E$           & $E$  &  $\Delta E$ & $E$ & $\Delta E$& $E$ & $\Delta E$ \\
    \midrule
	\endfirsthead
	\caption{(\emph{continued})} \\
	\toprule
    Key & State & NIST & \multicolumn{2}{c}{MBPT} &  \multicolumn{2}{c}{MCDF} & \multicolumn{2}{c}{Gu} & \multicolumn{2}{c}{Vilkas}\\
\cmidrule(r){4-5}\cmidrule(r){6-7}\cmidrule(r){8-9}\cmidrule(r){10-11}
&       &   $E$  &  $E$                       &  $\Delta E$           & $E$  &  $\Delta E$ & $E$ & $\Delta E$& $E$ & $\Delta E$ \\
    \midrule
    \endhead
    \midrule
	\endfoot
	\bottomrule
	\endlastfoot
1   & $2s^{2}\,2p^{4}(^{3}P)~^{3}P_{2}$       &    0        & 0        &       &   0       &        &    0        &       &   0       &            \\
2   & $2s^{2}\,2p^{4}(^{1}S)~^{1}S_{0}$       &    208080   & 207772   & -308  &   207830  &  -250  &    207878   &  -202 &   207956  &   -124     \\
3   & $2s^{2}\,2p^{4}(^{3}P)~^{3}P_{1}$       &    863620   & 863686   & 66    &   863215  &  -405  &    864158   &  538  &   864147  &   527      \\
4   & $2s^{2}\,2p^{4}(^{1}D)~^{1}D_{2}$       &    976560   & 976768   & 208   &   976432  &  -128  &    977254   &  694  &   977113  &   553      \\
5   & $2s^{2}\,2p^{4}(^{3}P)~^{3}P_{0}$       &    1920420  & 1921492  & 1072  &   1920667 &  247   &    1922438  &  2018 &   1922041 &   1621     \\
6   & $2s~^{2}S\,2p^{5}~^{3}P_{2}^{\circ}$    &    2380360  & 2379704  & -656  &   2380355 &  -5    &    2381134  &  774  &   2382023 &   1663     \\
7   & $2s~^{2}S\,2p^{5}~^{3}P_{1}^{\circ}$    &    2654560  & 2654158  & -402  &   2655091 &  531   &    2655814  &  1254 &   2656837 &   2277     \\
8   & $2s~^{2}S\,2p^{5}~^{3}P_{0}^{\circ}$    &    3289690  & 3289601  & -89   &   3289830 &  140   &    3291540  &  1850 &   3292303 &   2613     \\
\end{longtable}

\newpage
\linespread{1}
\tiny
\setlength{\tabcolsep}{5pt}
\begin{longtable}{lllllcccccccc}
	\caption{\label{table4} Comparison of wavelengths  ($\lambda$ in \AA) and transition rates ($A$ in $s^{-1}$)  for some transitions among the n = 2 states of O-like Mo. NIST -- the experimental wavelengths from the NIST ASD~\citep{Kramida.2018.V.p}; MBPT -- the present MBPT results; MCDF -- the present MCDF results.}\\
	\toprule
	$i$ & $j$  & Lower level & Upper level & \multicolumn{3}{c}{$\lambda$}  & \multicolumn{2}{c}{ Percentage difference for  $\lambda$ }& \multicolumn{2}{c}{$A$}&\multicolumn{2}{c}{Percentage difference for  $A$}\\
	\cmidrule(r){5-7}\cmidrule(r){8-9}\cmidrule(r){10-11}\cmidrule(r){12-12}
	 & & & & NIST & MBPT & MCDF &  NIST \& MBPT & NIST\& MCDF & MBPT & MCDF& MBPT\& MCDF\\
	\midrule
	\endfirsthead
	\caption{(\emph{continued})} \\
	\toprule

\midrule
	\endhead
	\midrule
	\endfoot
	\bottomrule
	\endlastfoot
3	&9	& $2s^{2}\,2p^{4}(^{3}P)~^{3}P_{1}$       &           	 $2s~^{2}S\,2p^{5}~^{1}P_{1}^{\circ}$  &  37.483 & 37.5089 &  37.4921 & 0.07 &  0.02 &  4.121E+10 &  4.009E+10 &  -3 \\
1	&7	& $2s^{2}\,2p^{4}(^{3}P)~^{3}P_{2}$       &           	 $2s~^{2}S\,2p^{5}~^{3}P_{1}^{\circ}$  &  37.661 & 37.6767 &  37.6635 & 0.04 &  0.01 &  3.046E+11 &  2.972E+11 &  -2 \\
7	&10 &$2s~^{2}S\,2p^{5}~^{3}P_{1}^{\circ}$     &          	   $2p^{6}~^{1}S_{0}$                    &  38.187 & 38.2035 &  38.2074 & 0.04 &  0.05 &  5.076E+11 &  4.917E+11 &  -3 \\
4	&9	& $2s^{2}\,2p^{4}(^{1}D)~^{1}D_{2}$       &           	 $2s~^{2}S\,2p^{5}~^{1}P_{1}^{\circ}$  &  39.161 & 39.1703 &  39.1541 & 0.02 & -0.02 &  4.447E+11 &  4.346E+11 &  -2 \\
2	&7	& $2s^{2}\,2p^{4}(^{1}S)~^{1}S_{0}$       &           	 $2s~^{2}S\,2p^{5}~^{3}P_{1}^{\circ}$  &  40.866 & 40.8766 &  40.8620 & 0.03 & -0.01 &  9.258E+10 &  9.073E+10 &  -2 \\
3	&8	& $2s^{2}\,2p^{4}(^{3}P)~^{3}P_{1}$       &           	 $2s~^{2}S\,2p^{5}~^{3}P_{0}^{\circ}$  &  41.221 & 41.2216 &  41.2097 & 0.00 & -0.03 &  3.267E+11 &  3.183E+11 &  -3 \\
1	&6	& $2s^{2}\,2p^{4}(^{3}P)~^{3}P_{2}$       &           	 $2s~^{2}S\,2p^{5}~^{3}P_{2}^{\circ}$  &  42.014 & 42.0220 &  42.0105 & 0.02 & -0.01 &  1.720E+11 &  1.677E+11 &  -3 \\
3	&7	& $2s^{2}\,2p^{4}(^{3}P)~^{3}P_{1}$       &           	 $2s~^{2}S\,2p^{5}~^{3}P_{1}^{\circ}$  &  55.842 & 55.8512 &  55.8074 & 0.02 & -0.06 &  1.985E+10 &  1.940E+10 &  -2 \\
\end{longtable}

\end{document}